\documentclass[twocolumn,aps,showpacs,preprintnumbers,amsmath,amssymb, superscriptaddress]{revtex4-1}
\pdfoutput=1

\usepackage{bm}
\usepackage{graphicx}
\usepackage{color,xspace}

\begin{document}

\title{Observation of an electron band above the Fermi level in FeTe$_{0.55}$Se$_{0.45}$ from \emph{in-situ} surface doping}

\author{P. Zhang}
\affiliation{Beijing National Laboratory for Condensed Matter Physics, and Institute of Physics, Chinese Academy of Sciences, Beijing 100190, China}
\author{P. Richard}\email{p.richard@iphy.ac.cn}
\affiliation{Beijing National Laboratory for Condensed Matter Physics, and Institute of Physics, Chinese Academy of Sciences, Beijing 100190, China}
\affiliation{Collaborative Innovation Center of Quantum Matter, Beijing, China}
\author{N. Xu}
\affiliation{Beijing National Laboratory for Condensed Matter Physics, and Institute of Physics, Chinese Academy of Sciences, Beijing 100190, China}
\affiliation{Paul Scherrer Institut, Swiss Light Source, CH-5232 Villigen PSI, Switzerland}
\author{Y.-M. Xu}
\affiliation{Materials Sciences Division, Lawrence Berkeley National Laboratory, Berkeley, California 94720, USA}
\author{J. Ma}
\affiliation{Beijing National Laboratory for Condensed Matter Physics, and Institute of Physics, Chinese Academy of Sciences, Beijing 100190, China}
\author{T. Qian}
\affiliation{Beijing National Laboratory for Condensed Matter Physics, and Institute of Physics, Chinese Academy of Sciences, Beijing 100190, China}
\author{A. V. Fedorov}
\affiliation{Advanced Light Source, Lawrence Berkeley National Laboratory, Berkeley, California 94720, USA}
\author{J. D. Denlinger}
\affiliation{Advanced Light Source, Lawrence Berkeley National Laboratory, Berkeley, California 94720, USA}
\author{G. D. Gu}
\affiliation{Condensed Matter Physics and Materials Science Department, Brookhaven National Laboratory, Upton, New York 11973, USA}
\author{H. Ding}\email{dingh@iphy.ac.cn}
\affiliation{Beijing National Laboratory for Condensed Matter Physics, and Institute of Physics, Chinese Academy of Sciences, Beijing 100190, China}
\affiliation{Collaborative Innovation Center of Quantum Matter, Beijing, China}

\date{\today}

\begin{abstract}
We used \emph{in-situ} potassium (K) evaporation to dope the surface of the iron-based superconductor FeTe$_{0.55}$Se$_{0.45}$. The systematic study of the bands near the Fermi level confirms that electrons are doped into the system, allowing us to tune the Fermi level of this material and to access otherwise unoccupied electronic states. In particular, we observe an electron band located above the Fermi level before doping that shares similarities with a small three-dimensional pocket observed in the cousin, heavily-electron-doped KFe$_{2-x}$Se$_2$ compound.
\end{abstract}



\maketitle

The recent discovery of strong coupling superconductivity in Fe-chalcogenides without holelike Fermi surface (FS) pocket at the Brillouin zone (BZ) center \cite{Qian_PRL2011,XPWang_KFeSe_Gap,D_MouPRL2011, Y_Zhang_NatureMat2011,D_LiuNCOMM2012,S_TanNMAT12,S_HeNMAT2013,PengPRL112} changed drastically the common knowledge on Fe-based superconductors \cite{RichardRoPP2011}. In contrast to the Fe-pnictide superconductors, these new materials exhibit a small electron pocket at the BZ center that has been attributed to the hybridization of Fe$3d$ and Se$4p_z$ orbitals \cite{SCWang_KFeSe}. Such hybridization is believed to be responsible for the large $J_3$ exchange constant \cite{F_Ma_PRL102} observed in XFe$_2$Se$_2$ (X = K, Tl, Rb) as compared to the Fe-pnictides \cite{M_Wang_NCOM}, and thus to tune the magnetic properties of its parent compound. Moreover, there is increasing evidence that fluctuations of the local antiferromagnetic exchange interactions ($J_1-J_2-J_3$ model) may be responsible for Cooper pairing in Fe-based superconductors \cite{HuJP_SR2012}. Interestingly, neutron scattering experiments on the 11-family of Fe-chalcogenide superconductors also suggest a large $J_3$ coupling constant \cite{LipscombePRL2011}. However, no clear electron band has been reported so far at the BZ center for this latter family of Fe-chalcogenide superconductors. 

\begin{figure}[!t]
\begin{center}
\includegraphics[width=0.48\textwidth]{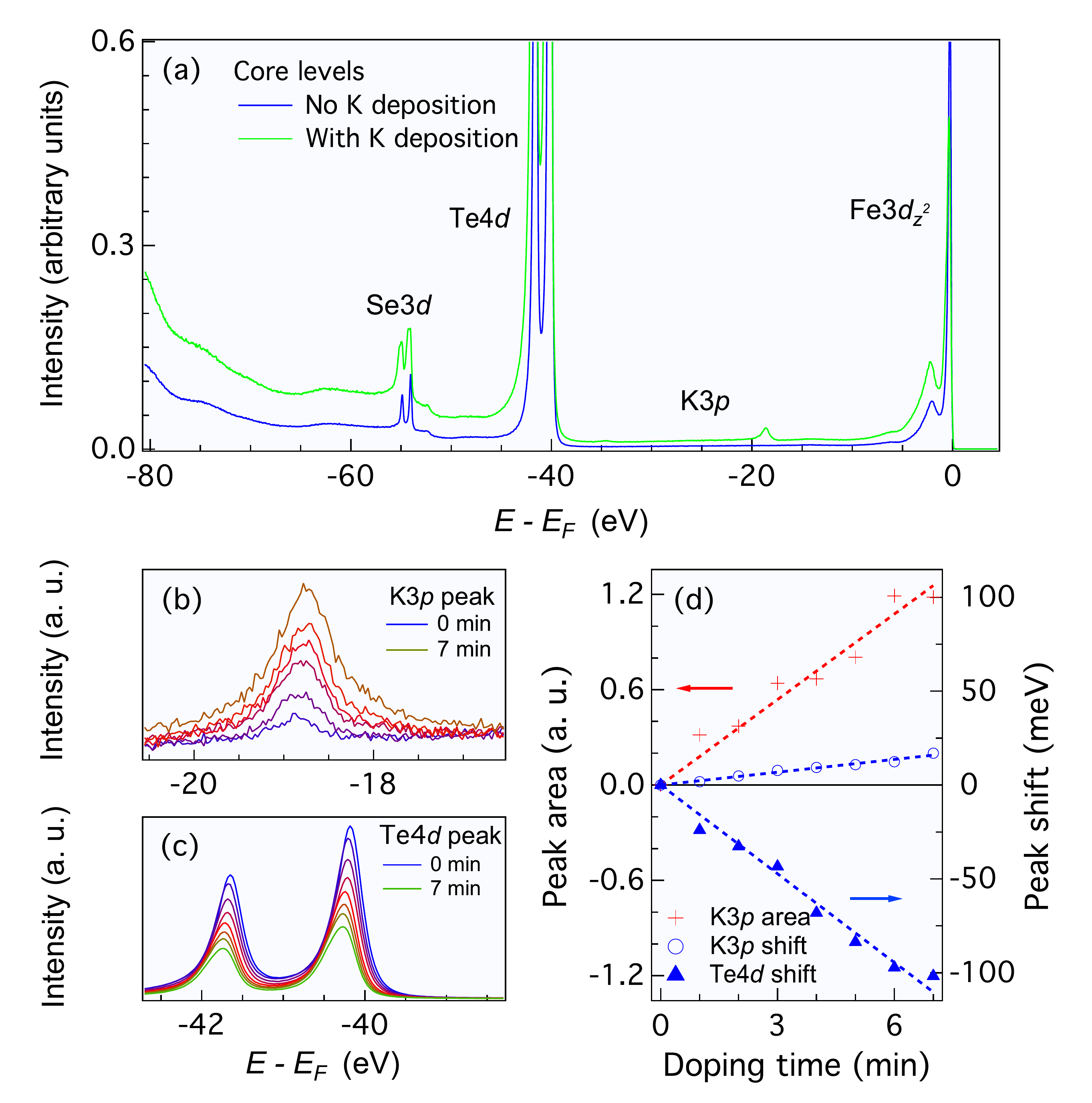}
\end{center}
 \caption{\label{corelevel} Color online. (\emph{a}) Core levels of FeTe$_{0.55}$Se$_{0.45}$ before (blue) and after (green) K surface doping, recorded with 105 eV photons. (\emph{b}) Evolution of the K$3p$ core level with K evaporation time. (\emph{c}) Evolution of the Te$4d$ core level with K evaporation time. (\emph{d}) Intensity (red) and peak shift (blue circle) of the K$3p$ core level as a function of K surface doping. The peak positions are fitted with a Gaussian and a linear background. The blue triangle represent the average peak shift of the Te$4d$ core levels}
\end{figure}

\begin{figure*}[!htb]
\begin{center}
\includegraphics[width=\textwidth]{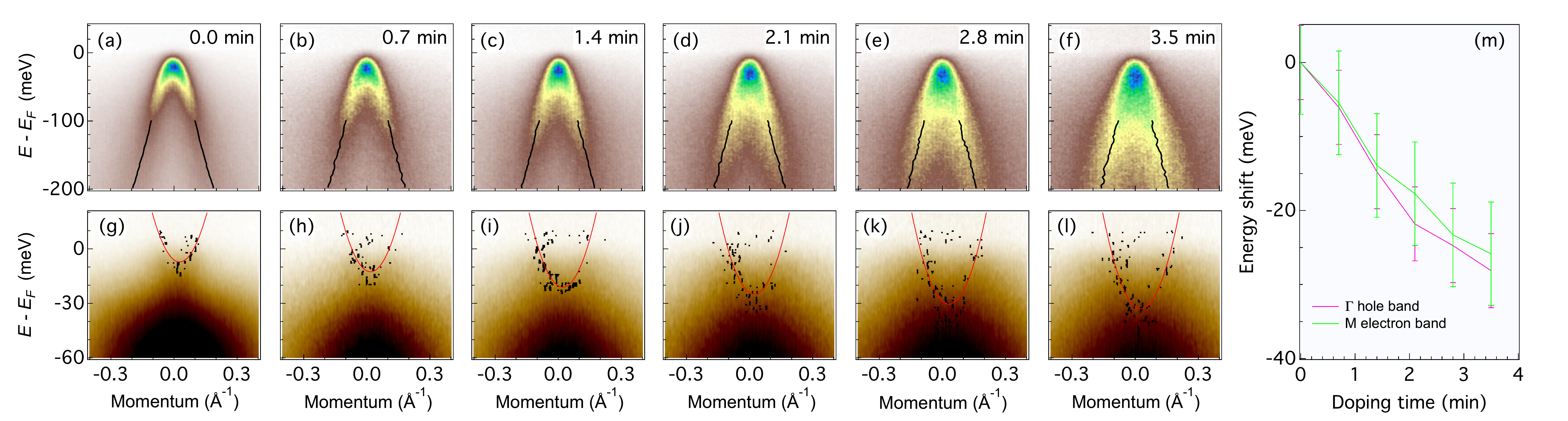}
\end{center}
 \caption{\label{shift} Color online. (\emph{a}) - (\emph{f}) ARPES intensity plots of the hole band at the $\Gamma$ point, recorded with 60 eV photons. The black lines are lorentzian fit results of the MDCs in the [-200, -100] meV range. (\emph{g}) - (\emph{l}) ARPES intensity plots of the electron band at the M point. The red curve is a parabolic fit for the electron band dispersion. (\emph{m}) Comparison of the the $\Gamma$ hole band shift derived from (\emph{a}) - (\emph{f}), and the electron band shift from (\emph{g}) -(\emph{l}). }
\end{figure*}

Angle-resolved photoemission spectroscopy (ARPES) is a powerful tool to resolve the single-particle electronic structure of crystalline materials, directly in the momentum space. However, the cut-off imposed by the Fermi-Dirac function limits ARPES to the observation of the occupied states and only to a very narrow energy range above the Fermi level ($E_F$). In this Letter, we use an \emph{in-situ} technique of K evaporation that has proved efficient to tune the chemical potential of topological insulators \cite{Hasan_TI1, Hasan_TI2} and cuprates \cite{Hossain_NP,Fournier_NPHYS6} without affecting the lattice parameters, and apply it to FeTe$_{0.55}$Se$_{0.45}$, a Fe-based superconductor. Our measurements reveal the existence at the BZ center of an electron band very near $E_F$, suggesting an electronic structure very similar to that of the 122-ferrochalcogenides.


Large single-crystals of FeTe$_{0.55}$Se$_{0.45}$ with high quality were grown using the self-flux method, and their $T_c$ was determined to be 14 K from magnetization measurements. ARPES measurements were performed at the Advanced Light Source, beamlines BL12 and BL4, using a VG-Scienta electron analyzer. The light used at the Advanced Light Source was linearly polarized in directions parallel or perpendicular to the analyzer slit. The angular resolution was set to 0.2$^{\circ}$. Clean surfaces for the ARPES measurements were obtained by cleaving the samples \emph{in situ} in a working vacuum better than 5 $\times$ 10$^{-11}$ Torr. All measurements were done between 11 K and 16 K. In the text we label the momentum values with respect to the 1 Fe/unit cell BZ. The K source used for evaporation is made of a SAES K dispenser, which is widely used to produce ultra pure K thin films. In the experiments, the largest coverage is less than one monolayer, as proved by the absence of splitting in the K$3p$ core level \cite{k3psplit}. 

In Fig. \ref{corelevel}\emph{a} we display the core level spectra of FeTe$_{0.55}$Se$_{0.45}$ before and after K surface doping. In addition to the Fe$3d$ states near $E_F$, we identify strong peaks associated with the Te$4d$ and Se$3d$ core levels. The most obvious difference between the two curves is an extra feature appearing around 19 eV after K doping, which corresponds to the K$3p$ core level and confirms that K atoms are deposited on the sample surface. In Fig. \ref{corelevel}\emph{b}, we show a zoom of the K$3p$ peak with increasing doping time. As expected, the peak intensity increases with doping, indicating a higher K content. In fact, the peak area and the peak shift given in Fig. \ref{corelevel}\emph{d} scale linearly with time (or doping) during the 7 minutes doping experiment. At least for the small doping doses used in our study, the coverage of K atoms is thus quite uniform. The Te$4d$ core levels, displayed in Fig. \ref{corelevel}\emph{c}, are also modified with K doping. In particular, as reported in Fig. \ref{corelevel}\emph{d}, the peaks shifts towards the high binding energies, suggesting an electron doping. We caution that the analysis of the core levels is always tricky, especially for adsorbate atoms, since the peak shift are not only the result of chemical potential shift, but also of the change in valency, in the Madelung potential, and in the core-carrier screening \cite{Hufner}. In fact, the shift of the adsorbate K core levels towards low binding energies can be predicted by applying a Born-Haber cycle within the equivalent-core approximation \cite{Pirug_SS163}. A more systematic approach to evaluate the chemical potential shift is to investigate directly the electronic dispersion of the low-energy sates.

\begin{figure*}[!htb]
\begin{center}
\includegraphics[width=\textwidth]{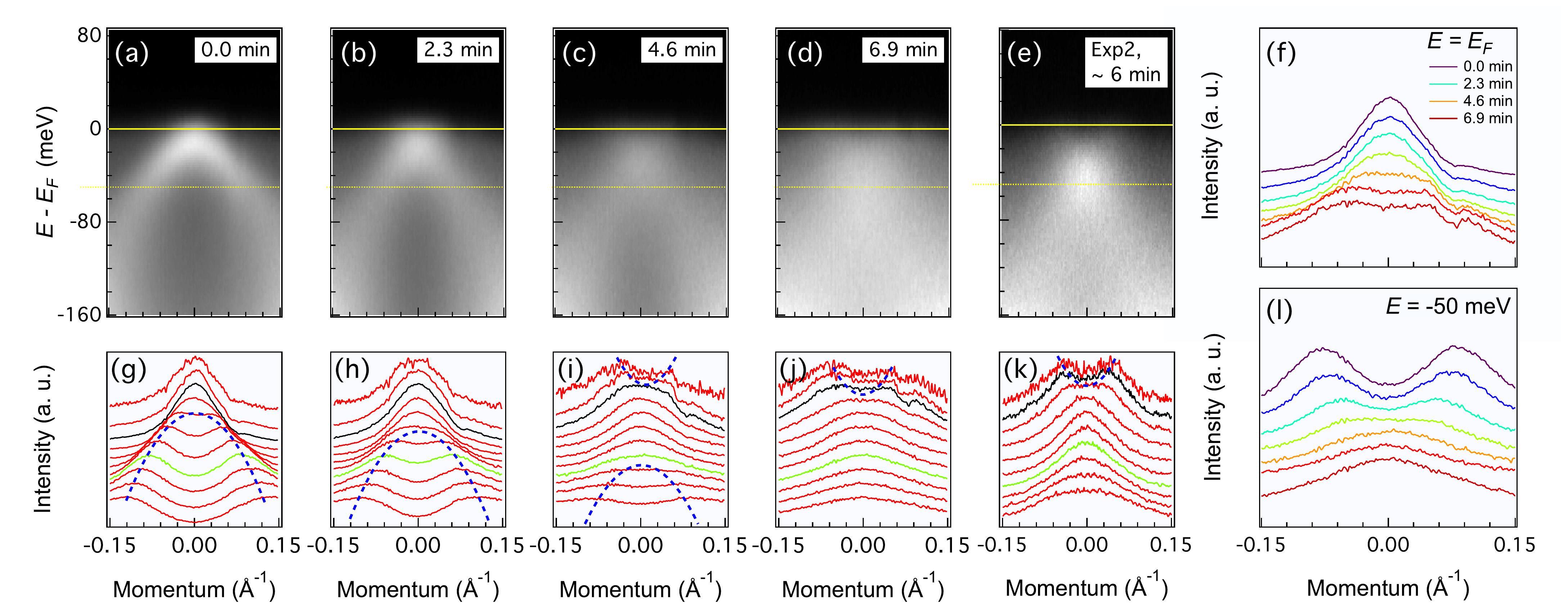} 
\end{center}
 \caption{\label{eb} Color online. (\emph{a}) - (\emph{d}) ARPES intensity plots around at $\Gamma$ with doping time, recorded with 30 eV photons. (\emph{e}) Similar experiment made with another sample (\emph{g}) - (\emph{k}) Corresponding MDC plots. The MDCs in black refer to $E_F$ while the ones in green have been measured 50 meV below $E_F$. The dashed lines are guides to the eye for the band dispersion. (\emph{f}) MDCs at $E_F$ with doping time. \emph{(l)} MDCs 50 meV below $E_F$ as a function of doping time.}
\end{figure*}

We are now going to confirm that electrons are introduced in the system \emph{via} K evaporation. In Figs. \ref{shift}(\emph{a}) - \ref{shift}(\emph{f}), we display the electron band structure at the $\Gamma$ point as a function of time (doping), using a $\pi$ configuration of polarization. According to the matrix element effect \cite{xpwang}, the hole band observed is the even composition of the Fe$3d_{xz}/d_{yz}$ orbitals. The top of this band is very difficult to determine precisely because it is located very close to the $E_F$ cutoff, as reported earlier \cite{MiaoPRB,Lubashevsky_NPHYS6,F_ChenPRB81,Starowicz_JPCM25}. However, we can track its evolution by tracing the band dispersion away from $E_F$, as obtained from fitting the momentum distribution curves (MDCs). Assuming a rigid band shift model, which is a good first approximation in the Fe-based superconductors \cite{Neupane_PRB2011},  we can deduce the chemical potential shift for a given evaporation time $t$ by shifting the electronic dispersion to match the one before evaporation ($t=0$, Fig. \ref{shift}(\emph{a})). The result, given by the pink curve in Fig. \ref{shift}(\emph{m}), shows a linear relationship, suggesting that the chemical potential is raised up due to the introduction of charge carriers.

Similar conclusion is also drawn from the band structure at the M point, as shown in Figs. \ref{shift}(\emph{g}) - \ref{shift}(\emph{l}). Although the band dispersion at the M point is not as clear as the one near the zone center, we can still track the band dispersion from the peak positions in the MDCs, here defined as locations of intensity larger than 0.98 in the MDCs renormalized to 1. The band dispersion can then be approximated by fitting a parabola through the clouds of points. As shown in Fig. \ref{shift}(\emph{m}) using green symbols, the bottom of the electron band shifts down, indicating and upward shift of the chemical potential. 

Interestingly, the comparison of the near-$E_F$ band shifts in Fig. \ref{shift}(\emph{m}) indicate a common shift for the bands very near $E_F$. The information ARPES can provide on the unoccupied states is directly limited by the Fermi-Dirac distribution function. With \emph{in-situ} doping, we are not only able to study the band shifts, but also to partly access electronic states that are unoccupied before doping. In Fig. \ref{eb}(\emph{a})-\ref{eb}(\emph{d}), we show the ARPES intensity plots of FeTe$_{0.55}$Se$_{0.45}$ around the $\Gamma$ point at different stages of K surface doping, which have been recorded using a $\sigma$ configuration of light polarization, with an extra polarization component along the $z$ axis. In this configuration, bands with an odd symmetry or with a $z$ component are visible. In addition, we show in Figs. \ref{eb}(\emph{g})-\ref{eb}(\emph{j}) the corresponding MDC plots. As expected, the intensity plots and the MDC plots show clearly that the holelike band is pushed towards higher binding energy as the K concentration increases. Moreover, these plots reveal an electron-like dispersion that is located above $E_F$ before the K evaporation. This has been confirmed in another experiment, for which the intensity plot and the MDC plot after 6 minutes evaporation are displayed in Figs. \ref{eb}(\emph{e}) and \ref{eb}(\emph{k}), respectively. In Figs. \ref{eb}(\emph{j}) and \ref{eb}(\emph{k}), this band clearly crosses $E_F$. To confirm that this band is not an artifact, we show in Figs. \ref{eb}(\emph{f}) and \ref{eb}(\emph{l}) the MDCs recorded as a function of doping at $E_F$ and 50 meV below $E_F$, respectively. Our observation is compatible with the observation of spectral weight above $E_F$ observed in a previous ARPES study of FeTe$_{0.6}$Se$_{0.4}$ \cite{Okazaki_SR4}.


Interestingly, a similar electron band was observed in the 122-chalcogenide XFe$_2$Se$_2$ (X = K, Tl, Rb) \cite{SCWang_KFeSe, XP_WangEPL2012,Y_Zhang_NatureMat2011}, which has the same Fe(Se,Te) layer as the material studied here but is heavily electron-doped. In the 122-chalcogenides, this small electron pocket has a three-dimensional character and originates mainly from the Se$4p_z$ orbital hybridized with the Fe$3d$ states. It is quite natural to assume that the small electron band evidenced in our study shares the same origin. As a result, our investigation shows that besides different chemical potentials, XFe$_2$Se$_2$ and Fe(Se,Te) have qualitatively the same electronic band structure. The proximity of the small electron pocket to $E_F$ suggests that the chalcogenide atoms may play a more important role in the hopping terms than their pnictide counterparts, in agreement with neutron scattering experiments showing a larger $J_3$ exchange constant in the ferrochalcogenides than in the ferropnictides \cite{LipscombePRL2011}. Our results may have direct consequences on the superconductivity in these materials, either by opening additional scattering channels due to the presence of an additional band near $E_F$ or by modulating the local exchange interactions, and in particular the $J_3$ term, believed to be responsible for the modulation in the ferrochalcogenides of the pairing symmetry away from the simple $\cos(k_x)\cos(k_y)$ global gap function \cite{XPWang_KFeSe_Gap,MiaoPRB,HuJP_SR2012}.

In conclusion, \emph{in-situ} K evaporation can be used to dope the surface of Fe-based superconductors and tune their chemical potential. In particular, we studied  the near-$E_F$ dispersion of FeTe$_{0.55}$Se$_{0.45}$. We showed the existence of an electron-like pocket at the zone center quite similar to a small three-dimensional FS observed previously in the cousin, heavily-electron-doped KFe$_{2-x}$Se$_2$ compound. Our study provides a way to explore the unoccupied states of the Fe-based superconductors. 

We acknowledge X. Dai and Z. Fang for useful discussions. This work was supported by grants MOST (2010CB923000,  2011CBA001000, 2011CBA00102, 2012CB821403 and 2013CB921703) and NSFC (11004232, 11034011/A0402, 11234014 and 11274362) from China. The Advanced Light Source is supported by the Director, Office of Science, Office of Basic Energy Sciences, of the U.S. Department of Energy under Contract No. DE-AC02-05CH11231.

\bibliography{biblio_doping}

\end{document}